\documentstyle[11pt,epsfig,epsf]{article} 
\def\baselinestretch{1.3}
\newcommand{\ba}{\begin{array}}
\newcommand{\ea}{\end{array}}
\newcommand{\bd}{\begin{displaymath}}
\newcommand{\ed}{\end{displaymath}}
\newcommand{\be}{\begin{equation}}
\newcommand{\ee}{\end{equation}}
\newcommand{\bea}{\begin{eqnarray}}
\newcommand{\eea}{\end{eqnarray}}

\def\q2 {q^2}

\def \gev {GeV}

\begin{document}
\begin{flushright}
{\large MRI-P-001102\\ 
hep-ph/0011240}\\
\end{flushright}
\begin{center}
{\Large\bf  Probing the violation of equivalence principle \\ 
at a muon storage ring via neutrino oscillation}\\[14mm]
{\sf Anindya Datta}  

{\em Harish-Chandra
Research Institute,\\
Chhatnag Road, Jhusi, Allahabad - 211 019, India}

{E-mail: anindya@mri.ernet.in}
\end{center}

\vskip 20pt
\begin{abstract}
  
  We examine the possible tests of violation of the gravitational
  equivalence principle (VEP) at a muon storage ring via neutrino
  oscillation experiments. If the gravitational interactions of the
  neutrinos are not diagonal in the flavour basis and the
  gravitational interaction eigenstates have different couplings to
  the gravitational field, this leads to the neutrino oscillation. If
  one starts with $\mu ^+$ beam then appearance of $\tau ^\pm$, $e ^+$
  and $\mu ^-$ in the final state are the signals for neutrino
  oscillation. We have estimated the number of $\mu ^-$ events in this
  scenario in $\nu _\mu -N$ deep inelastic scattering.  Final state
  lepton energy distribution can be used to distinguish the VEP
  scenario from the others. A large area of VEP parameter space can be
  explored at a future muon storage ring facility with moderate beam
  energy.

\end{abstract}

\noindent
\section {General Strategy}

The oscillation among the different neutrino flavours is now a well
accepted solution to the solar neutrino problem
\cite{neutrino_osc,solar_neutrino}.  Recent results from the
SuperKamiokande (SK) experiment \cite{atmos_neutrino} at Japan on
atmospheric neutrinos also support this proposition. The most popular
explanation behind the neutrino oscillation is that neutrinos have
non-degenerate masses and the mass eigenstates are not the same as 
gauge interaction eigenstates. Oscillation probability from one
flavour to another is proportional to $\Delta m^2$, where $\Delta
m^2$ is the difference between the square of the physical masses.
Lots of work have already been done in this direction
\cite{other_works}.

An alternative to this ``mass-mixing (MM) solution'' was proposed long
back by Gasperini and independently by Halprin and Leung \cite{VEP}.
Their idea was that gravitational interaction of the neutrinos may not
be diagonal in the flavour basis. Thus the gravitational interaction
eigenstates are different from the weak interaction eigenstates.  Now
if these gravitational interaction eigenstates couple to the
gravitational potential with different strength, neutrino oscillation
can take place.  This mechanism does not require neutrinos to have
non-zero masses.

The principle of equivalence has been the cornerstone of the general
theory of relativity.  Validity of this principle has been tested to a
high precision for macroscopic bodies.  But this has not been tested
experimentally in the microscopic and quantum regime. So it is prudent
to keep an open mind towards this issue and try to prove/disprove the
validity of equivalence principle in the current and future
experiments.

For a two flavour oscillation picture\footnote{we will stick to this
in the present analysis due to simplicity. But this is sufficient to
illuminate the underlying principle.}, the
most general expression for the transition probability is given by,

\be 
P_{i \leftrightarrow j} = \sin ^2 2\theta \;\sin^2 \left(\frac{\pi
    L}{\lambda}\right)
\label{osc_prob}
\ee
L, is the base-line length (distance which neutrinos traverse from the
source to the detector). $\theta$ is the usual mixing angle.

Expressions for $\lambda$ for the two cases (MM and VEP) are, 
\bea
\lambda &=& \frac{4 \pi E_\nu}{\Delta m^2}~~(MM)  \nonumber \\
&=& \frac{2 \pi}{E_{\nu} \phi \Delta f}~~(VEP) 
\eea 
In the above
equations, $E_\nu$ is the neutrino energy.  $\Delta f ~(\equiv f_1 -
f_2)$ represents the difference between the coupling strengths of the two
gravitational eigenstates with gravitational potential $\phi$ and it
quantifies the violation of equivalence principle in the neutrino
sector. The basic difference between the two cases, on
the dependence of $E_\nu$, is evident. So their predictions can be
different altogether. This alternative has attracted many people and a
lot of work has been done \cite{vep_other}.  Oscillation probability
induced by VEP is,
\be 
P = \sin ^2 2 \theta \; \sin ^2 \left(2.538
\times 10^{18}\; L E_\nu \Delta f \phi \right)
\ee 
Here, L is
expressed in kilometres and $E_\nu$ in $\gev$.  We will assume that
$\phi$ remains unchanged over the neutrino path. But, this is not very
crucial for our purpose, we will parametrise VEP by the product
$\phi\;\Delta f \;(\equiv \Delta F)$. Some recent work \cite{arc},
analysing the SK solar neutrino data, gives the best fit value for
this parameter. This comes out to be as small as 10$^{-24}$ for
maximal mixing.

In this letter, we will try to explore the possible signals of VEP at
a muon storage ring.  Here, we assume the sample design, for muon
production, capture, cooling, acceleration and storage as given in
ref.  \cite{geer}. Number of available muons directed to the neutrino
detector of per year is $10^{20}$ (one can have $\mu^+$ or $\mu^-$) .
For the purpose of illustration, we start with $\mu ^+$ beam.  These
positively charged muons will produce $\sim$ $10^{20}$ $\bar \nu
_\mu$s and same number of $\nu _e$s.  Muon anti-neutrinos (or electron
neutrinos) thus produced, traverse a distance L before colliding on a
fixed target. For a sufficiently energetic initial muon beam,
neutrino-nucleon (in the target material) interaction is in the
deeply-inelastic regime. Now, on their way to the detector, if some
$\bar \nu_\mu$s are oscillated to $\bar \nu_e$ or $\bar \nu_\tau$s, in
the final state we may observe $\tau$s or $e$s with the same sign with
the initial muon beam, via neutrino nucleon charged current
interaction.  The ${\nu}_e$s (also coming from $\mu$ decay) may
transform to ${\nu}_\mu$s or ${\nu}_\tau$s, which can be the source of
$\mu ^-$s or $\tau ^-$s. We will call them wrong sign $\mu$s or
$\tau$s, as they have opposite charge to the initial $\mu$-beam. The
un-oscillated $\bar \nu_\mu$ (${\nu}_e$)s can only give rise to $\mu
^+$ ($e^-$)s.  So appearance of $\tau$ leptons of either sign and
negatively charged muons or positively charged electrons are
definitely the signal for neutrino oscillation at a muon storage ring
with positively charged muon-beam.  These issues have been studied in
detail earlier \cite{suku} in the context of MM scenario. Here we will
mainly concentrate on the results for $\mu ^-$ appearance in the final
state.  This means our focus is on the $\nu_e \leftrightarrow \nu_\mu$
oscillation due to VEP. We will see in the following, that $\mu ^-$
event rate in MM scenario is negligibly small with presently SK
allowed values for $\Delta m^2_{e\mu}$ and $\sin^2 2\theta$.  So
appearance of considerable number of $\mu^-$ events at a muon storage
ring, is a signature which goes against the MM solution to neutrino
oscillation problem.  So the charge identification for the muons is
necessary for the study of neutrino oscillation physics. This seemed
to be achievable according to the ref.\cite{geer}.  VEP solution does
not fit very well to the $\nu _\mu - \nu_\tau$ oscillation data from
SK. So we will not calculate the $\tau ^+$ appearence rate in the 
following.

For lower values of  energy of the initial muon beam, $\tau ^-$
cross-section is slightly lower than the $\mu^-$ case due to the phase
space suppression.  Furthermore, for tau in the final state, above
cross-section has to be multiplied by appropriate branching ratio of
$\tau$. Generally one prong hadronic or leptonic decay channel is the
cleanest way to detect a $\tau$.  Former has a branching ratio of 45
\% whereas the later has of 17 \%.

Number of $\mu ^-$ events  coming from neutrino-nucleon DIS
can be obtained by folding the charged current cross-section by
oscillation probability, neutrino flux and finally by the total number
of nucleons, $N_n$, present in the fixed target. This last quantity
depends on the amount of the target material present.

\be N_{\mu } = N_{n} \int \sigma(\nu_\mu + N
\rightarrow \mu^-  + X)\; \frac{dN_{\nu}}{dE_{\nu_e}} \;P(\nu_e 
\leftrightarrow \nu_\mu)\; dE_{\nu_e} 
\label{no_of_evts}
\ee

We present our results for 1 kT of the target
material\footnote{This corresponds to $ N_n = 6.023 \times 10^{32}$}.
For any other detector size these results can easily be scaled.
The charged-current neutrino-nucleon
cross-section is calculated assuming almost the same number of protons
and neutrons, present in the target material. The expression for
the same is not given here but can be obtained in several places
\cite{geer,suku}.  Our results are based on a simple parton level
monte-carlo event generator.  We have not incorporated any
detector effects.  Even incorporating these, the essence of our
analysis will remain the same.  We have used CTEQ4LQ parametrisations
\cite{cteq} for the parton distribution functions to estimate the
above cross-section.

Neutrino flux depends on the number of neutrinos,
initially present. At a muon storage ring this number is
equal to the number of muons present in the beam. Neutrino flux also
depends on energy/angular distribution of the neutrinos coming from
$\mu $ decay.  We will elaborate more on this in the following.

The area of the target and the base-line length, L, define a cone with
half angle $\theta_d$, which can be written as $R_d = L\;\theta_d$.
($R_d$ is the radius of the detector if we assume it has a circular
cross-section.)  Thus the choice of detector size for a long-baseline
experiment must differ with that of a short base-line experiment and
also depends on the energy of the muon beam.  If the muon-beam energy
is increased, neutrinos coming from muon decay, become more and more
boosted in the muon direction.  This effectively increases the
luminosity of neutrino beam. A higher effective luminosity of the
incident neutrino beam can be achieved with a sufficiently high energy
beam and/or by decreasing the effective area of the target (keeping
the amount of target material fixed).  In our event generator, we
only consider those $\nu$s, for which $\theta _{\mu,\nu} < \theta_d$.
($\theta _{\mu,\nu}$ is the angle between the decaying muon beam and
the neutrino.)  Of course this is fully determined by the muon decay
kinematics.  In this analysis we will assume a fixed detector size
with $R_d \sim 6 ~mts.$. Target area corresponding to this radius are
similar to that of proposed ICANOE detector \cite{icanoe}.

\section {Discussion of the Results}

\begin{figure}[ht]
\centerline{ 
\epsfxsize=10cm\epsfysize=7.0cm
                     \epsfbox{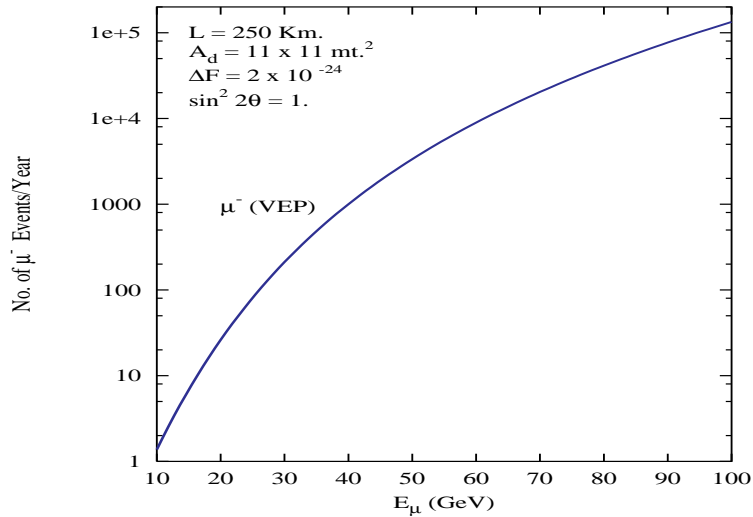}
}

\caption{{\it Number of $\mu ^-$  events coming from 
 $\nu_\mu $-N 
    DIS at a muon storage ring  with
    $\Delta F$ =2 $\times 10^{-24}$ and $\sin ^2 2\theta$ = 1. $A_d$
is the area of the target. }}
\label{fig:fig1}
\end{figure}

In Fig. \ref{fig:fig1}, we present variation of the number of wrong
sign muon events coming from ${\nu}_\mu$-N DIS experiments for VEP
scenario. Number of tau events, from ${\nu}_\tau$-N scattering will
almost be the same as this. We have chosen $\Delta F$ = 2$\times 10 ^{-24}$
and $\sin ^2 2 \theta$ as 1.  Choice of
these parameter values are from ref.  \cite{arc}. They have
used the SK solar neutrino data, to calculate the best fit values of
$\Delta F$ and $\sin ^2 2 \theta$. The only other bounds on these two
parameters come from the LSND and E776 experiment at BNL
\cite{utpal_prl}. Ref. \cite{arc} constrains $\Delta F$ responsible
for the $\nu_e \leftrightarrow \nu_x ~(x \equiv \mu, \tau ~or
~stetrile)$ oscillation. The other one \cite{utpal_prl} constrains the same 
parameters for $\nu_e \leftrightarrow \nu_\mu$ oscillation.

We will not present here the $\tau ^+$ event rate from $\bar \nu _\tau
- N$ scattering. VEP solution to atmospheric $\nu_\mu - \nu_\tau$
oscillation is disfavoured by the the SK data
\cite{lipari,Fogli,Sobel}. In ref. \cite{Fogli,Sobel}, a $\chi ^2$
analysis has been done with SK atmospheric neutrino data, assuming a
general power law dependence of oscillation probability on the
neutrino energy ($\sim E^n$).  Minimum $\chi ^2$ is obtained for $n =
-1$ {\em i.e.} for the MM scenario.

But it should be borne in mind that along with this indirect evidence
($\chi ^2$ analysis in the ref. \cite{Fogli,Sobel}) against the VEP
solution to the atmospheric $\nu_\mu - \nu_\tau$ oscillation, one also
should look for some direct evidences to disprove such propositions.
Neutrino factory with a huge and precisely known neutrino flux is an
ideal place for this.  $\bar \nu_\mu \leftrightarrow \bar \nu_\tau$
oscillation due to VEP and subsequent $\bar \nu_\tau - N$ scattering
can produce $\tau ^+$ events. The search for $\tau$ appearence in
final state will be one of the major goals of neutrino oscillation
experiments. So we would not have to pay some extra price to look for
$\tau^+$ signal which can can arise from VEP in neutrino sector.
    
We have checked that event rates remain almost unchanged for two
baseline lengths (First one, 250 Km., corresponds to a baseline from
Kamioka to KEK and the second one, 740 Km., corresponds to a baseline
from CERN to Gran Sasso or from Fermilab to Soudan experiment.) over
the energy range we used in fig.  \ref{fig:fig1}.  For the values of
neutrino energy and $\Delta F$ we have used, oscillation probability
grows quadratically with L. At the same time, for a fixed target area
and size, effective luminosity of colliding neutrinos decreases with
L, due to the decrement of the detection cone subtended by the target.
These two effects compensate each other, for the detector area we
have chosen.  We checked that for a sufficiently large target area,
and/or sufficiently high muon beam energy, when almost all the
neutrinos coming from muon decay can be intercepted by the target,
(i.e. when neutrino flux becomes independent to L) the event rate
indeed grows quadratically with L.

Number of $\mu ^-$ events in MM scenario, when calculated using SK
allowed values for mass$^2$ difference and mixing angle ($\Delta m^2_{e \mu}
= 2 \times 10^{-5} eV^2, ~\sin^2 2\theta = 1$. These values are for MSW
solution. Vacuum solution gives $\Delta m^2_{e\mu} \sim 10^{-10} eV^2$
\cite{talk_gonzales}), comes out to be small compared to the VEP
scenario over the range of muon energies we are interested and is beyond 
the scale used in fig.\ref{fig:fig1}. The other source of $\mu ^-$ is
from the $\tau ^-$ (produced in $\nu_\tau - N$ interaction) decay. But
this rate is further suppressed by $\tau \rightarrow \mu$ branching
ratio.  Thus the appearance of $\mu ^-$ in final state (starting with
$\mu ^+$ beam), is a definite signature of neutrino oscillation not
originated by MM scenario. Appearance of $\mu ^-$ in the final state
may also be accounted by the R-parity violating $\mu$-decay and/or
$\nu - N$ interaction \cite{our_rp}\footnote{R-parity violating
theories allow lepton flavour violation, thus $\mu ^-$ or $\tau$
appearance in such a theory can also be explained without flavour 
oscillation.}.

We confine our discussion to the appearance of wrong sign muons in the
final state. Wrong sign $\tau$ event rate will be somewhat lower (due
to phase-space suppression) than this if difference of couplings
between $\nu_\mu$ and $\nu_\tau$ to the gravitational potential is at
the same ballpark of $\nu_\mu$ and $\nu_e$. Apart from ref.
\cite{arc}, the only bound on the $\Delta F$, coming from two
terrestrial experiments (LSND and E776) are not consistent with each
other. And these values are order of magnitude greater than the best
fit value obtained in ref. \cite{arc}.  Upper bound coming from the
E776 experiment is around 3$\times 10 ^{-21}$ for $\sin ^2 2 \theta$ =
1. On the other hand LSND results predicts a lower bound on $\Delta F
~(\sim 5 \times 10 ^{-18}$) for the same value of the mixing angle.
The SK situation is a bit different from the other two cases
(In terrestrial experiments, neutrinos are mainly affected by the
earth's gravitational field. In the former, neutrinos traverse the
sun-earth distance. Gravitational fields of sun and near by stars
along with the earth have to be considered in such a case.).  We want
to stress that even with such a small value of the above parameter,
number of $\mu ^-$ events in VEP scenario will be much larger than
that of MM scenario.  In the following we will explore the possible
region in $\Delta F - \sin ^2 2 \theta$ space for which a perceptible
signal can be obtained such that we can exclude that region at a
certain confidence level.

\begin{figure}[ht]
  \centerline{ \epsfxsize=10cm\epsfysize=7.0cm \epsfbox{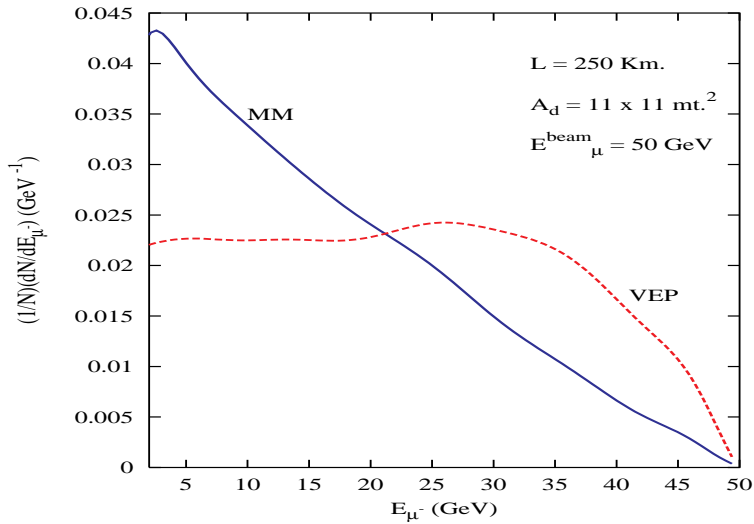} }

\caption{{\it Normalised $\mu ^-$ energy distribution at a 
    50 GeV muon storage ring. Solid curve is for MM scenario and the
    dashed curve is for VEP scenario $A_d$ is the target area.}}
\label{fig:fig2}
\end{figure}

Before we go into the exclusion contours, we want to point out another
distinguishing feature of the VEP scenario, which is also an artefact
of the energy dependence of oscillation probability.  We look at the
normalised energy distribution of the $\mu ^-$ in the final state. In
MM scenario lepton energy distribution falls of rapidly in contrast to
the VEP scenario. This is reflected in fig.  \ref{fig:fig2}. If the
VEP parameter $\Delta F$ is so small that event rate is comparable
with event rate in the MM scenario \footnote{One can see from previous
  discussion, $\mu^-$ event rate in MM scenario is negligibly small
  compared to the same in VEP scenario with the values of parameters
  we have used.  But having larger amount of target material and at
  the same time smaller value of $\Delta F$ can make these two
  comparable with each other.}, energy distribution of the final state
lepton can be used to discriminate between these two scenarios. So one
can have an idea about the energy dependence of the oscillation
probability from the lepton energy distribution.  If one concentrates
on the $\mu ^-$ appearance rate, then MM scenario is outnumbered by
VEP scenario.  Strong constraint can be put on this scenario from the
wrong sign muon search.  We have already mentioned about $\mu ^-$ or
$\tau ^+$ appearance due to the R-parity violating interaction at a
muon storage ring. Wrong sign muon or tau events can be accounted
without the oscillation phenomena in such a kind of theories. Here
also the lepton energy distribution is expected to be different from
MM scenario.

Until now, we have used the best fit values for $\Delta F$ and $\sin ^2 2
\theta$ as given in ref. \cite{arc}, but the main purpose of this
letter is to point out that even a tiny violation of equivalence
principle can be detected at a muon storage ring facility and can be
distinguished from the MM scenario quite efficiently.  In the
following we like to show, in the $\Delta F - \sin ^2 2\theta$ plane,
the region which can be excluded at 95 \% C.L.

From our previous discussions, it is clear that $\mu^-$ appearance
search is the best probe for VEP at a neutrino factory.  Firstly,
$\mu^-$ event rate is higher than the $\tau^-$ event rate.  And with
experimentally allowed values of MM parameters, one cannot have
 $\mu^-$ signal from MM scenario.  We will present the exclusion
contours from $\mu^-$ appearance search for
two values of beam energy.

We will not discuss here in details the possible backgrounds for such
signals. These issues have been discussed in ref. \cite{suku_cut}.  We
have applied the same kind of cuts to reduce the background.  \bea p_T
^{\mu ^-} &>& 2 ~GeV, \nonumber \\ \Delta R_{\mu^- X} &>& 0.4 \eea
$\Delta R_{\mu^- X}$ is the isolation between the $\mu ^-$ and the DIS
hadronic products.  After applying the above set of cuts the signal
efficiency comes out to be 70 \%.  The other source of wrong sign muon
is from the decay of the wrong sign taus' coming from the ${ \nu}_\tau
- N$ scattering. These ${\nu}_\tau$s come from ${ \nu}_e
\leftrightarrow {\nu}_\tau$ oscillation. So this is also due to
flavour oscillation and add to our signal. But we confine ourselves to
two flavour oscillation only and do not add the numbers of wrong sign
muons coming from tau decay.
\begin{figure}[ht]
\centerline{ 
\epsfxsize=10cm\epsfysize=7.0cm
                     \epsfbox{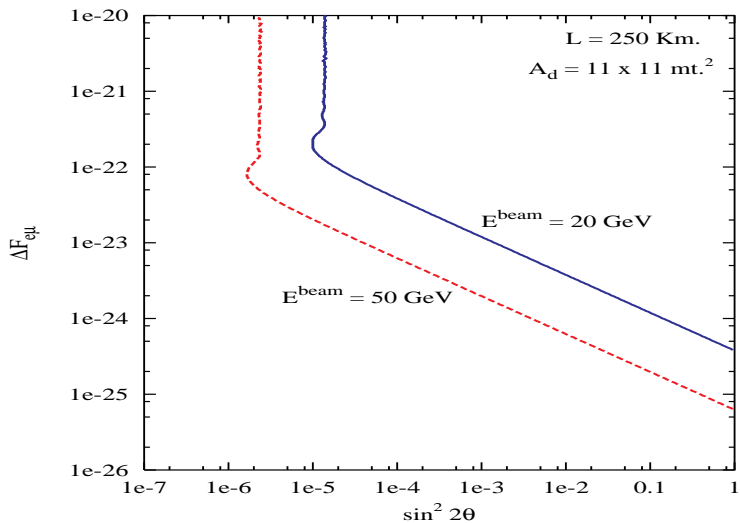}
}

\caption{{\it 95 \% exclusion contour in $\Delta F_{e\mu} - \sin ^2 2 \theta$ plane
for 50 and 20 GeV muon storage ring from wrong sign muon signal. 
}}
\label{fig:fig3}
\end{figure}

 The exclusion contour depicts the fact
that for small values of $\Delta F$ one need maximal mixing between
the flavours.  For higher values of VEP parameter very small mixing
would sufficient to produce enough number of signal events. For
$\Delta F$ less than $10 ^{-23}$ , $\sin ^2 2\theta$ and $\Delta F$
are correlated in a linear fashion. In this region, oscillation
wave-length $\lambda$ (see eqn. \ref{osc_prob}) is much larger than
the base-line length, L. For larger values of $\Delta F$, L is
comparable with $\lambda$. Here, $\sin^2(\Delta F \,E_\nu \,L)$
attains its maximum value. This explains the correlation between the
two oscillation parameters near the `knees' of the contours. For even
higher values of $\Delta F$, $L >> \lambda$ and the $\sin^2(\frac{\pi
L}{\lambda})$ can be approximated by $\frac{1}{2}$. This causes the
sharp rise of the contour almost independent of mixing angle.

A different phenomenon which also cause the neutrino oscillation with
the same neutrino energy dependence of oscillation probability, is the
violation of special relativity (VSR) \cite{VSR}. One can parametrised
the oscillation probability in the same way. So the above analysis
is completely applicable to VSR also. The energy distribution and
exclusion contours derived above would be the same for 
the later.

\section{Conclusions}

To summarise, we have shown that at a muon storage ring with moderate
energy, neutrino flavour oscillation due to VEP can be probed.  The
$\mu ^-$ appearance event rate due to $\nu_\mu -N$ DIS has been
calculated within RG improved parton model. We compared the VEP
results with the MM results and have shown that, the former
will give overwhelmingly large number of wrong sign muon events than
the later. In VEP model, the number of $\mu ^-$ events increases with
beam energy. The energy distributions of the final state muon have also
been compared for two different scenario.  If the violation of
equivalence principle is so small that it would produce number of
events comparable with that of mass mixing model, then energy
distribution of the final state lepton can be used as a discriminator.
Finally we have shown the 95 \% C.L. exclusion contours in the $\Delta
F_{e\mu} - \sin ^2 2\theta$ plane for two different muon beam energy.

{\large \bf Acknowledgement}
The author acknowledges useful discussions and suggestions regarding this
work with A.  Raychaudhuri, B. Mukhopadhyaya and U. Sarkar.

\setcounter{footnote}{0}

\def\baselinestretch{1.8}

\newcommand{\plb}[3]{{Phys. Lett.} {\bf B#1} (#3) #2}                  %
\newcommand{\prl}[3]{Phys. Rev. Lett. {\bf #1} (#3) #2}        %
\newcommand{\rmp}[3]{Rev. Mod.  Phys. {\bf #1} (#3) #2}             %
\newcommand{\prep}[3]{Phys. Rep. {\bf #1} (#3) #2}                     %
\newcommand{\rpp}[3]{Rep. Prog. Phys. {\bf #1} (#3) #2}             %
\newcommand{\prd}[3]{Phys. Rev. {\bf D#1} (#3) #2}                    %
\newcommand{\prc}[3]{{Phys. Rev.}{\bf C#1} (#3) #2}  
\newcommand{\np}[3]{Nucl. Phys. {\bf B#1} (#3) #2}                     %
\newcommand{\npbps}[3]{Nucl. Phys. B (Proc. Suppl.) 
           {\bf #1} (#3) #2}                                           %
\newcommand{\sci}[3]{Science {\bf #1} (#3) #2}                 %
\newcommand{\zp}[3]{Z.~Phys. C{\bf#1} (#3) #2}                 %
\newcommand{\mpla}[3]{Mod. Phys. Lett. {\bf A#1} (#3) #2}             %
 \newcommand{\apj}[3]{ Astrophys. J.\/ {\bf #1} (#3) #2}       %
\newcommand{\astropp}[3]{Astropart. Phys. {\bf #1} (#3) #2}            %
\newcommand{\ib}[3]{{ ibid.\/} {\bf #1} (#3) #2}                    %
\newcommand{\nat}[3]{Nature (London) {\bf #1} (#3) #2}         %
 \newcommand{\app}[3]{{ Acta Phys. Polon.   B\/}{\bf #1} (#3) #2}%
\newcommand{\nuovocim}[3]{Nuovo Cim. {\bf C#1} (#3) #2}         %
\newcommand{\yadfiz}[4]{Yad. Fiz. {\bf #1} (#3) #2;             %
Sov. J. Nucl.  Phys. {\bf #1} #3 (#4)]}               %
\newcommand{\jetp}[6]{{Zh. Eksp. Teor. Fiz.\/} {\bf #1} (#3) #2;
           {JETP } {\bf #4} (#6) #5}%
\newcommand{\philt}[3]{Phil. Trans. Roy. Soc. London A {\bf #1} #2  
        (#3)}                                                          %
\newcommand{\hepph}[1]{(electronic archive:     hep--ph/#1)}           %
\newcommand{\hepex}[1]{(electronic archive:     hep--ex/#1)}           %
\newcommand{\astro}[1]{(electronic archive:     astro--ph/#1)}         %

\end{document}